\documentclass[twocolumn]{aastex63}
\usepackage{lineno}

\newcommand{\lsun}{\mbox{L}_\odot}
\newcommand{\rsun}{\mbox{R}_\odot}
\newcommand{\msun}{\mbox{M}_\odot}
\newcommand{\lacc}{L_{\rm acc}}
\newcommand{\macc}{\dot{M}_{\rm acc}}
\newcommand{\lstar}{L_\star}
\newcommand{\mstar}{M_\star}
\newcommand{\rstar}{R_\star}
\newcommand{\teff}{T_{\rm eff}}

\newcommand{\rhole}{R_{\rm hole}}
\newcommand{\rdisk}{R_{\rm disk}}

\definecolor{green}{rgb}{0.0, 0.5, 0.0}

\received{\today}
\revised{}
\accepted{}
\submitjournal{ApJ}

\shorttitle{Accretion variability of VW Cha}
\shortauthors{Zsidi et al.}
\graphicspath{{./}{figures/}}

\begin{document}

\title{Accretion variability of the multiple T Tauri system VW Cha}

\author[0000-0002-4612-5824]{Gabriella Zsidi}
\email{zsidi.gabriella@csfk.org}
  \affiliation{Konkoly Observatory, Research Centre for Astronomy and Earth Sciences, E\"otv\"os Lor\'and Research Network (ELKH), Konkoly-Thege Mikl\'os \'ut 15-17, 1121 Budapest, Hungary}
  \affiliation{ELTE E\"otv\"os Lor\'and University, Institute of Physics, P\'azm\'any P\'eter s\'et\'any 1/A, 1117 Budapest, Hungary}

\author[0000-0002-5261-6216]{Eleonora Fiorellino}
  \affiliation{Konkoly Observatory, Research Centre for Astronomy and Earth Sciences, E\"otv\"os Lor\'and Research Network (ELKH), Konkoly-Thege Mikl\'os \'ut 15-17, 1121 Budapest, Hungary}

\author[0000-0001-7157-6275]{\'Agnes K\'osp\'al}
  \affiliation{Konkoly Observatory, Research Centre for Astronomy and Earth Sciences, E\"otv\"os Lor\'and Research Network (ELKH), Konkoly-Thege Mikl\'os \'ut 15-17, 1121 Budapest, Hungary}
  \affiliation{Max Planck Institute for Astronomy, K\"onigstuhl 17, 69117 Heidelberg, Germany}
  \affiliation{ELTE E\"otv\"os Lor\'and University, Institute of Physics, P\'azm\'any P\'eter s\'et\'any 1/A, 1117 Budapest, Hungary}

\author[0000-0001-6015-646X]{P\'eter \'Abrah\'am}
  \affiliation{Konkoly Observatory, Research Centre for Astronomy and Earth Sciences, E\"otv\"os Lor\'and Research Network (ELKH), Konkoly-Thege Mikl\'os \'ut 15-17, 1121 Budapest, Hungary}
  \affiliation{ELTE E\"otv\"os Lor\'and University, Institute of Physics, P\'azm\'any P\'eter s\'et\'any 1/A, 1117 Budapest, Hungary}

\author[0000-0002-8585-4544]{Attila B\'odi}
  \affiliation{Konkoly Observatory, Research Centre for Astronomy and Earth Sciences, E\"otv\"os Lor\'and Research Network (ELKH), Konkoly-Thege Mikl\'os \'ut 15-17, 1121 Budapest, Hungary}
  \affiliation{MTA CSFK Lend\"ulet Near-Field Cosmology Research Group, 1121, Budapest, Konkoly Thege Mikl\'os \'ut 15-17, Hungary}

\author[0000-0003-3547-3783]{Gaitee Hussain}
  \affiliation{European Space Agency (ESA), European Space Research and Technology Centre (ESTEC), Keplerlaan 1, 2201 AZ Noordwijk,The Netherlands}

\author[0000-0003-3562-262X]{Carlo F. Manara}
  \affiliation{European Southern Observatory, Karl-Schwarzschild-Strasse 2, 85748 Garching bei München, Germany}

\author[0000-0001-5449-2467]{Andr\'as P\'al}
  \affiliation{Konkoly Observatory, Research Centre for Astronomy and Earth Sciences, E\"otv\"os Lor\'and Research Network (ELKH), Konkoly-Thege Mikl\'os \'ut 15-17, 1121 Budapest, Hungary}
  



\begin{abstract}

Classical T Tauri stars are low-mass objects, which are still accreting material from the surrounding circumstellar disk. The accretion process is essential in the formation of Sun-like stars and in setting the properties of the disk at the time when planet formation occurs. We constructed a complex dataset in order to examine the accretion process of VW Cha, a classical T Tauri multiple system with the aim of studying the physical origin of the photometric and spectroscopic variability of the system. The TESS Space Telescope observed VW~Cha between 2019 April 22 and June 19, and we complemented these data with contemporaneous ground-based $I_CJHK$ band photometric measurements. In addition, we obtained high-resolution optical spectra with the VLT/ESPRESSO and the 2.2\,m/FEROS instruments. Analyzing these data, we found that the TESS light curve shows photometric variations on timescales from minutes to weeks with a peak-to-peak amplitude of $\sim$0.8 mag. The near-infrared light curves follow the shape of the optical measurements, however, the peak-to-peak amplitudes are slightly increasing towards the longer wavelengths. We took spectra in both fainter and brighter photometric states of the system, allowing us to examine the origin of a photometric brightening event. Our results show that this brightening event can be explained by increased accretion. In addition, our spectroscopic data also suggest that the primary component of VW~Cha is a spectroscopic binary, as it was proposed in earlier works.

\end{abstract}



\section{Introduction} \label{sect:intro}

Classical T Tauri stars (CTTSs) are young low-mass objects ($<2-3$\,M$_{\odot}$), which are still surrounded by a circumstellar disk.
They host magnetic fields which are strong enough to truncate the circumstellar disk at a few stellar radii \citep{johns-krull1999}. According to the magnetospheric accretion model, the material that falls onto the star is channeled along the magnetic field lines \citep{hartmann2016}. This accretion process is inherently variable and it causes both photometric and spectroscopic variations \citep{herbst1994, hartmann2016}.

The accretion process is often accompanied by outflows as well in the form of winds or jets \citep{hartmann2016, frank2014}. These are considered to be driven by rotating magnetic fields anchored in the inner disk. The atomic and ionized components of the outflows can be traced in visual spectra by, e.g., blueshifted forbidden emission lines  such as [O~I], [S~II] or [N~II] \citep[][]{bally2016}, or blueshifted absorption components of permitted emission lines (e.g., the Balmer and Paschen series of hydrogen).

Accretion and outflows are not the only processes that are able to cause changes in T~Tauri systems and it has been proven that the optical-infrared variability of CTTSs arises from a combination of physical processes occurring at or close to the stellar surface \citep{cody2014}. Cool magnetic spots can appear on the stellar surface, and as they modulate the observed brightness of the star during its rotation \citep{bouvier1993}, they cause periodic variation, however, irregular fluctuation due to accretion complicates their recognition. At the same time, additional irregular variability can occur due to obscuration by the circumstellar matter \citep[e.g.][]{herbst1994}. 
The effect of all the above mentioned phenomena combined with the unsteady accretion \citep{bouvier2007} appear superimposed in the observations.

The majority of stars form in a binary or a multiple system. Their configuration ranges from close binaries with circumbinary disks to systems with wide companions and individual circumstellar disks. Their evolution is affected by the companion: close, eccentric binaries tend to produce pulsed accretion \citep{kospal2018}, however, wide companions might have only minor effect.
Examining multiple systems with high-resolution multi-technique campaigns combining spectroscopy and photometry would help understanding their evolution and the impact of the companion on the circumstellar disk, which might host planets, via studying the details of the accretion, outflows, or winds.

This study focuses on VW~Cha, which is a young low-mass classical T Tauri multiple system \citep{brandner1996,brandeker2001,nguyen2012} in the Chamaeleon~I star forming region at a distance of $d=194.5^{+10.6}_{-9.1}$\,pc \citep{bailerjones2021}. 
This source has already been studied in previous works, mainly as part of larger samples.
\cite{dae13} carried out a survey of multiple systems in the Chamaeleon~I region. They resolved the primary and the secondary components of VW~Cha, and their study resulted in 
$T^{\rm A}_{\rm eff} = 4060$\,K and  $T^{\rm B}_{\rm eff} = 3850$\,K, 
$L^{\rm A}_{\star} = 1.6\,\lsun$ and $L^{\rm B}_{\star} = 0.76\,\lsun$, 
$M^{\rm A}_\star = 0.75\,\msun$ and $M^{\rm B}_\star = 0.57\,\msun$, 
$R^{\rm A}_\star = 2.55\,\rsun$ and $R^{\rm B}_\star = 1.97\,\rsun$,
spectral types of K7 and M0,
and ages of 1.0 and 1.5~Myr for the A and B components, respectively.
The study provides a lower limit for the accretion luminosity and accretion rate of the primary with
$\log L^{\rm A}_{\rm acc} = -1.51\,\lsun$, $M^{\rm A}_{\rm acc} = 4.20\cdot 10^{-9}\,\msun/ \rm yr$, whereas \cite{man16a} reported $\log L_{\rm acc} = -0.78\,\lsun$, $\log M_{\rm acc} = -7.60\,\msun/ \rm yr$ and $A_V=1.9$\,mag for the system, both assuming a distance to Chamaeleon~I of 160\,pc. 
\cite{banzatti2015} reported a disk inclination of 44$^{\circ}$ based on the rovibrational band of CO near 4.7\,$\mu$m.

In this paper, we aim to study the accretion variability of the multiple system VW~Cha using ground-based and space photometry, and high-resolution spectroscopy. 
We describe our observations in Section \ref{sect:obs}, show our results in Section \ref{sect:results} and discuss them in Section \ref{sect:discussion}.

\begin{figure*}[ht!]
    \centering
    \includegraphics[width=\textwidth]{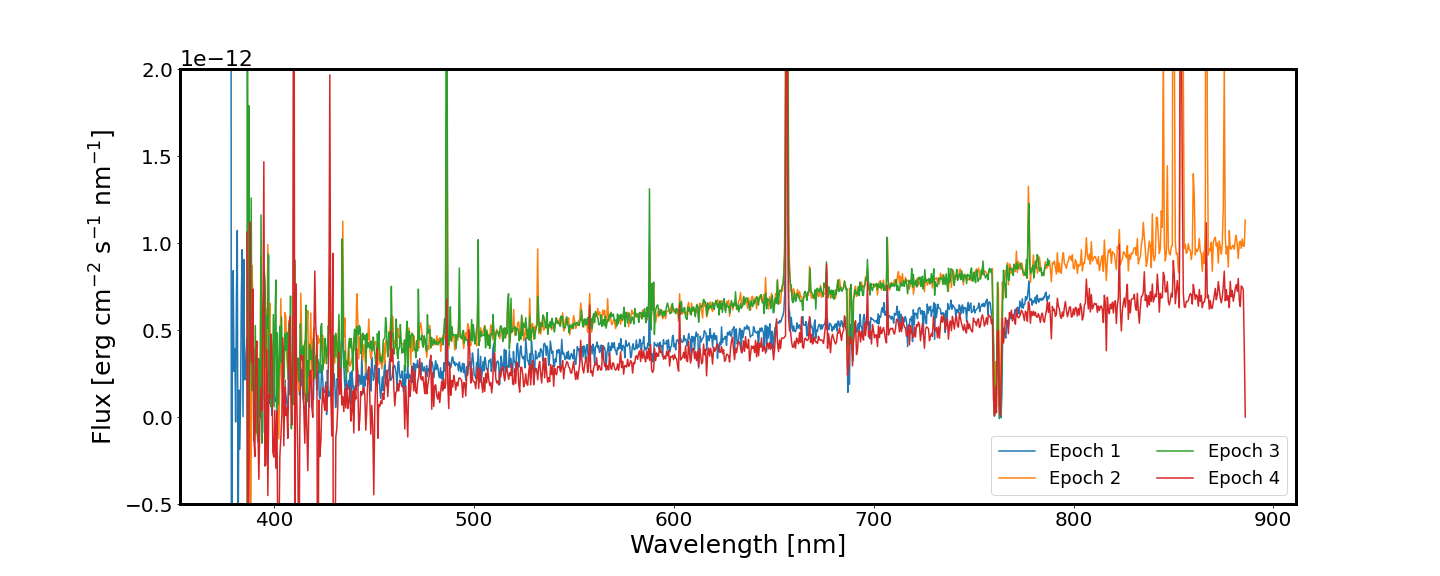}
    \caption{The flux calibrated and smoothed FEROS and ESPRESSO spectra. }
    \label{fig:smooth}
\end{figure*}

\section{Observations} \label{sect:obs}

The Transiting Exoplanet Survey Telescope (TESS) covered parts of the Chamaeleon~I star forming region by Sectors 11 and 12 in 2019. We designed a contemporaneous observing program including multi-filter ground-based photometric monitoring and high-resolution spectroscopic measurements in order to study the variability and the accretion process in T~Tauri systems.

TESS provided a 56-day-long light curve of VW~Cha with 30-minute cadence between 2019 April 22 and 2019 June 19. We carried out the data reduction via differential photometry, using the \texttt{ficonv} and \texttt{fiphot} tools of the FITSH package \citep{pal2012}. As a reference flux is needed for the process, we used the median of our own SMARTS $I_C$ band photometry, taken contemporaneously with the TESS observations.
More detailed description on the TESS data reduction and photometry process can be found in \citet{plachy2021} and \citet{pal2020}. We also inspected the most recent TESS light curves of VW~Cha because it was also covered by Sectors 38 and 39 in 2021. This resulted in an additional 58-day-long light curve with a more frequent, 10-minute cadence between 2021 April 28 and 2021 June 24.

We obtained ground-based photometric observations contemporaneously with the 2019 TESS light curve using the 1.3 m telescope (Cerro Tololo, Chile) operated by the SMARTS Consortium. The observations were performed with approximately nightly cadence in the $I_CJHK$ bands, covering the time interval between 2019 April 29 and June 10. In addition, in the first three nights $V$ and $R_C$ band measurements were also taken. We typically obtained 7 images in the $I_C$~band with an exposure time of 14\,s, and 5--7 frames in the near-infrared bands with exposure times of 12\,s (4\,s at the first two epochs) in the $J$ band and 4\,s in the $H$ and $K$~bands. The optical images were delivered to us by the SMARTS team after standard bias and flatfield corrections. For the $JHK$ images we performed dithering in order to enable bad pixel removal and sky subtraction. These steps were completed by using our custom IDL scripts. In the optical bands the photometric calibration was based on two comparison stars in the $6'\times6'$ field of view (2MASS~ J11072405-7741258 and 2MASS~J11081648-7744371) whose magnitudes were extracted from the APASS9 catalog \citep{henden2015} and transformed to the Johnson-Cousins system using the equations of \citet{jordi2006}. The near-infrared photometric calibration was based on the 2MASS magnitudes \citep{cutri2003} of a carefully selected comparison star, 2MASS~J11075699--7741558. Due to this procedure, although the $JHK$ observations were carried out in the CIT/CTIO $JHK$ filters, the final calibrated magnitudes quoted in Table~\ref{tab:photometry} are in the 2MASS photometric system.

We carried out high-resolution (R=140\,000) optical (380--788\,nm) spectroscopic observations with the ESPRESSO instrument mounted on the Very Large Telescope as part of a DDT proposal (Pr.Id.2103.C-5025, PI \'A. K\'osp\'al) on 2019 May 31 and June 8. 
We carried out the bias, dark, and flat field corrections and the wavelength calibration using Version 3.13.2 of the EsoReflex/ESPRESSO pipeline \citep{freudling2013}. The resulting 2D spectra were merged producing a rebinned 1D spectra.
We obtained two additional high-resolution (R=48\,000) optical (350--920\,nm) spectra on June 8 and June 11 with the FEROS instrument on the MPEG/ESO 2.2~m telescope in object-calib mode, where contemporaneous spectra of a ThAr lamp were recorded throughout the whole object exposure. 
The \textsc{ferospipe} pipeline in \textsc{python} is available for reducing data acquired with the FEROS instrument, detailed description of the pipeline can be found in \cite{brahm2017}. However, this pipeline was designed to precisely measure the radial velocity, and it calibrates only 25 of the available 33 \'echelle orders. The excluded orders are essential for our analysis as they cover the $\sim$673-823\,nm range including multiple accretion tracer lines, therefore, we modified the original pipeline in a way that we extended the wavelength coverage to all orders. More details on the modified pipeline is described in \cite{nagy2021}.

The flux calibration of all spectra was performed in two steps. First, we normalized the spectra by fitting the continuum. 
Then, we calculated the calibration coefficients using the ASAS-SN ($g$ band, $\lambda_{\textrm{eff}} = 470$~nm) and TESS ($I_C$ band, $\lambda_{\textrm{eff}} = 784$~nm) photometry, obtained contemporaneously with the ESPRESSO and the FEROS spectra.
Magnitudes were converted into fluxes by using the zero fluxes\footnote{\url{http://svo2.cab.inta-csic.es/theory/fps/}} at 470~nm and 784~nm, for $g$ and $I_C$ bands respectively.
We obtained the calibrated spectra by multiplying the normalized spectra by the calibration function, which was developed by linearly interpolating the calibration coefficients in $g$ and $I_C$ bands. 
The flux calibrated spectra are shown in Fig.~\ref{fig:smooth}.

\begin{deluxetable}{cccc}
\tablecaption{Log of the spectroscopic observations. \label{tab:spectra}}
\tablewidth{0pt}
\tablehead{
\colhead{Date (Epoch)} & \colhead{JD$-$2450000} & \colhead{Instrument} & \colhead{Seeing [$\arcsec$]}
}
\startdata
2019 May 31 (Ep 1) & 8635.49998 & ESPRESSO & 0.96 \\
2019 June 08 (Ep 2) & 8642.52353 & FEROS & 1.26 \\
2019 June 08 (Ep 3) & 8642.66358 & ESPRESSO & 0.47 \\
2019 June 11 (Ep 4) & 8645.57175 & FEROS & 1.70
\enddata
\end{deluxetable}

\begin{figure*}[t]
    \centering
    \includegraphics[width=\textwidth]{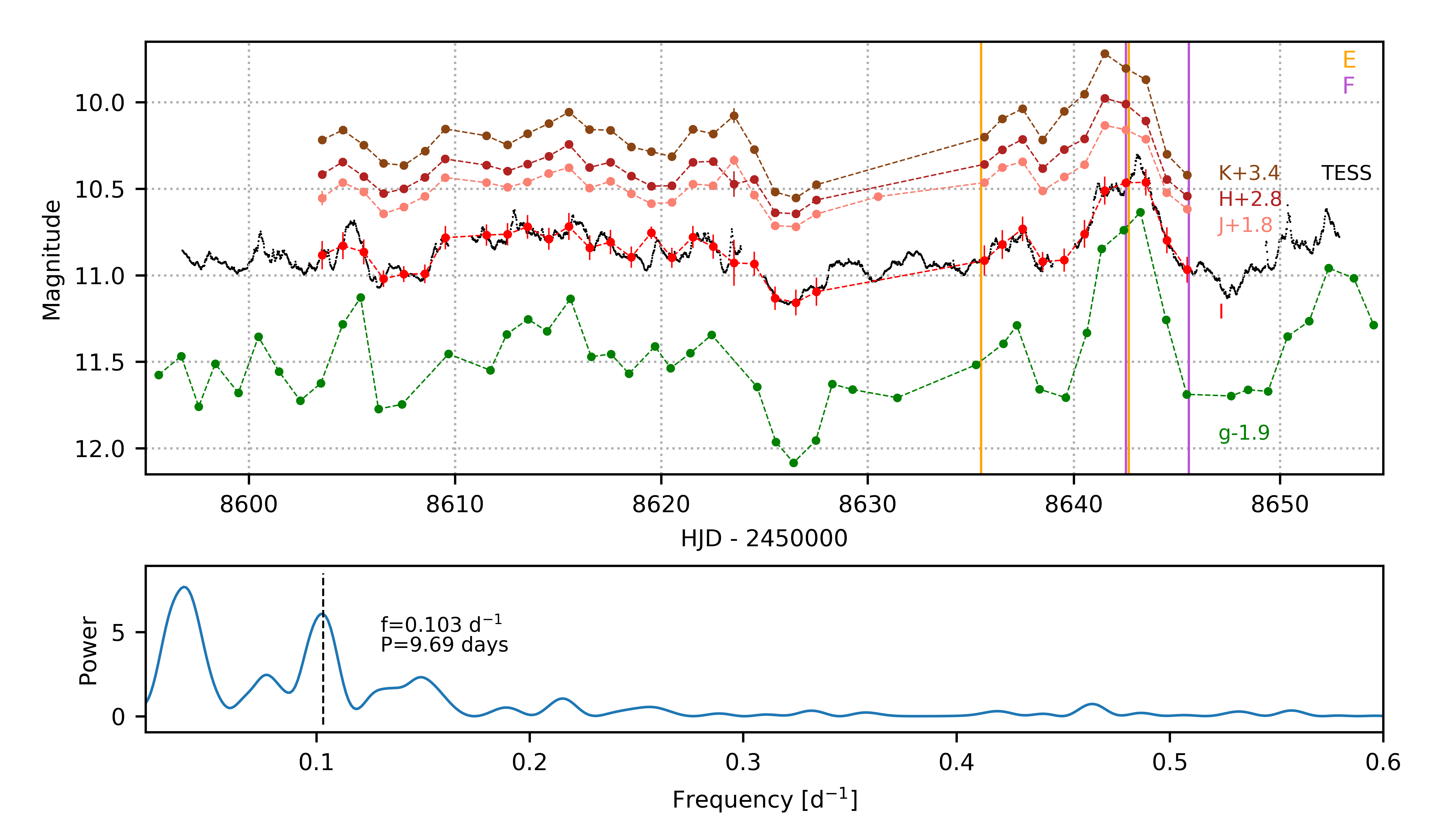}
    \caption{The light curves and the periodogram of VW Cha in 2019. The top panel shows our near-infrared SMARTS $JHK$ band light curves. We show our optical $I_C$ band SMARTS photometry with red circles, the TESS light curve with black dots, and the nightly averaged ASAS-SN $g$ band data with green circles. The $JHK$ and $g$ band light curves are shifted along the $y$ axis with the values indicated in the Figure. We marked the epochs of the spectroscopic observations with vertical orange (ESPRESSO) and purple (FEROS) lines. The bottom panel shows the Lomb-Scargle periodogram obtained from the TESS data.}
    \label{fig:lc_period}
\end{figure*}

\subsection{Our observations in the context of the multiplicity of VW~Cha}

VW~Cha is a known multiple T~Tauri system. It was identified as a binary by \cite{brandner1996} with 0$\farcs$72 separation and $\Delta Z = 0.25$\,mag brightness difference between the companion and the primary at 1\,$\mu$m. 
\cite{brandeker2001} reported an additional close companion to the secondary star with a separation of 0$\farcs$10 and suggested that VW~Cha is a physical triple system. 
They found that the primary increasingly dominates the emission at increasingly longer wavelengths and reported a brightness difference of ${\Delta}J=0.73$\,mag  and ${\Delta}K=1.32$\,mag between the primary (A) and the secondary (Ba+Bb) components. \citet{dae13} also detected this close companion of the secondary. 
\cite{correia2006} found a wide companion (C) at 16$\farcs$8. 
\cite{melo2003} suggested that the primary component (A) is a double line spectroscopic binary (SB2), and \citet{nguyen2012} also reported that the primary is a suspected SB2. We will examine further this possibility in Sect.~\ref{sect:sb2}.

In order to investigate which component contributes to the observed flux and the variability, we examined the Gaia EDR3 measurements. According to these data, the separation between the A and the B components is 0$\farcs$664 and the position angle is 185.778$^{\circ}$. The primary component is slightly brighter with Gaia magnitude G=12.40\,mag for A and G=12.77\,mag for B, i.e., 58.3\,\% of the total flux comes from the primary component and 41.7\,\% from the companion.
In our photometric observations, we did not resolve the primary (A) and the secondary (B) components.

 Our spectroscopic observations were made with fibre fed spectrographs. ESPRESSO has a 1$\farcs$0 aperture, which means that when the fibre is centered on the primary, the secondary at 0$\farcs$72 separation is excluded from the aperture, however, due to varying seeing conditions (see Table \ref{tab:spectra}), we might observe some contamination from the companion. The FEROS instrument has larger aperture (2$\farcs$0), which means that the secondary contaminates our observations. 
 Furthermore, if the primary (A) is an SB2, as suggested by \cite{nguyen2012} and \cite{melo2003}, both components would have been observed by the ESPRESSO and the FEROS spectrographs.
 In addition, according to \cite{dae13}, the probability of A component hosting a disk is 1.00, whereas the same probability of the B component is only 0.01. For this reason, we attribute the accretion related variability to the primary component.

\begin{figure*}[t]
    \centering
    \includegraphics[width=\textwidth]{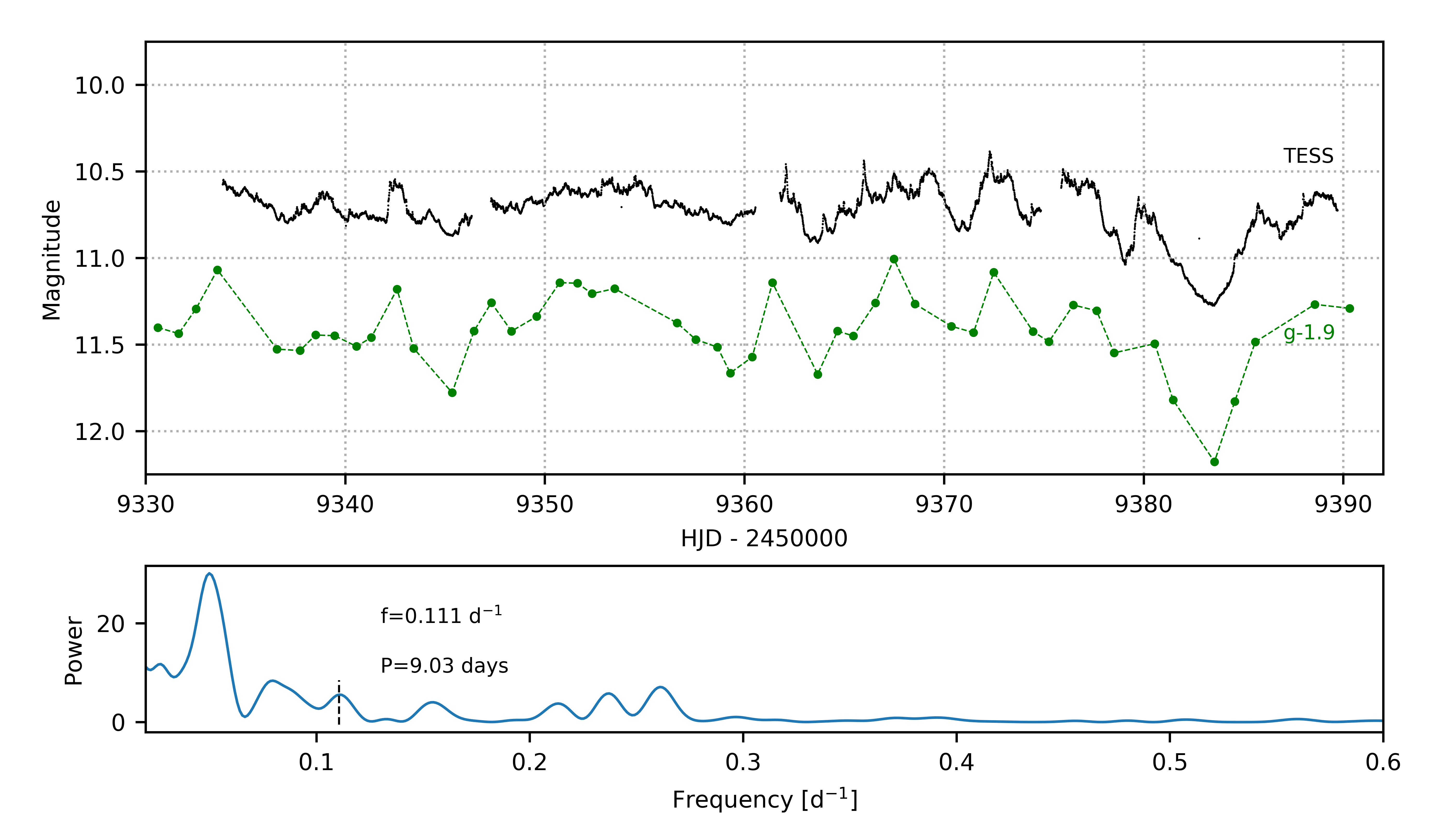}
    \caption{The light curves and periodogram from 2021. In the top panel, the black dots show the TESS observations from 2021 and the green circles indicate the nightly averaged ASAS-SN $g$ band data. The bottom panel shows the Lomb-Scargle periodogram obtained from the 2021 TESS data.}
    \label{fig:lc_period2021}
\end{figure*}

\section{Results} \label{sect:results}

\subsection{Light curves} \label{sect:light_curves}

The light curves of VW~Cha carry important information on the nature of the variability and its origin. 
The TESS light curve, displayed with black dots in Fig.~\ref{fig:lc_period}, reveals variations on various timescales with a peak-to-peak amplitude of $\sim$0.8\,mag. Small amplitude fluctuations appear on timescales of hours, and a brightening event occurs towards the end of our observing period (JD$\sim$2458642, i.e., around 2019 June 7), which lasts for a few days. 
In order to look for any periodic signals in the TESS light curve, we carried out a period analysis. We computed a Lomb-Scargle periodogram \citep{lomb1976, scargle1982}, which reveals a period at $P=9.69$\,days (bottom panel of Fig.~\ref{fig:lc_period}). We also found a peak around 26 days ($f=0.037$~d$^{-1}$), which suggests a longer-term oscillation in the light curve. No further peaks were found with higher frequencies outside of the indicated frequency range of Fig.~\ref{fig:lc_period}.
The 2021 TESS light curve (Fig.~\ref{fig:lc_period2021}) shows similar peak-to-peak amplitude ($\sim$0.9\,mag), however, the mean brightness of the system slightly increased. 
We calculated a Lomb-Scargle periodogram from the 2021 TESS light curve as well (Fig.~\ref{fig:lc_period2021}, bottom panel), which does not confirm the presence of the 9.69\,days period but shows a small peak at 9.03\,days instead. As the TESS photometry is not resolving the system, the results of the period analysis might be the superposition of the signals produced by all members of the system. 


Although the ground-based SMARTS observations have sparser cadence than the TESS measurements, they carry crucial information on the color variations during our observing period. 
The $I_C$ band measurements perfectly follow the variations seen in the TESS light curve with similar peak-to-peak amplitude. This is expected, as TESS offers broad band photometry, which is centered at the $I_C$ band \citep[see Fig. 1 in][]{ricker2015}. 
The shape of the near-infrared $J$, $H$ and $K$ band light curves also follow the shapes of the optical light curves with slightly increasing amplitudes towards the longer wavelengths, i.e., $\sim$0.6\,mag peak-to-peak variations in the $J$ band and $\sim$0.8\,mag variations in the $K$ band. 

\begin{figure}[t]
    \centering
    \includegraphics[height=0.5\textwidth, angle=90]{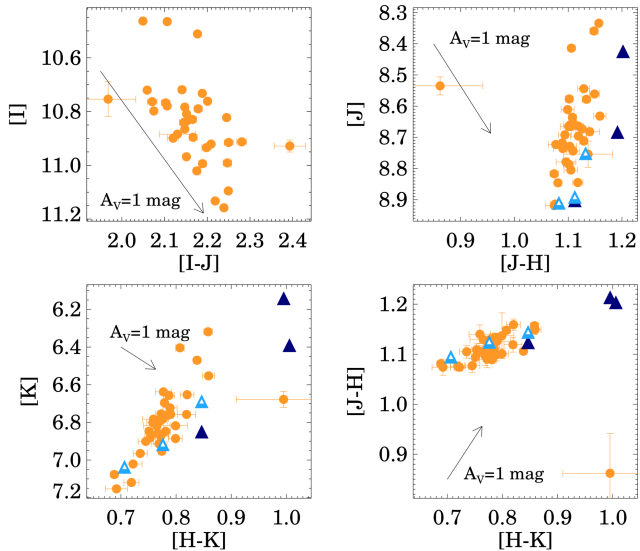}
    \caption{The color-magnitude and color-color diagrams obtained from the SMARTS data. We drew the extinction corresponding to $A_V$=1 mag with black arrow. We indicated the values from the disk model of Fig. 25. from \cite{carpenter2001} with triangles. The open symbols represent mass accretion rate of $10^{-8.5}$\,M$_{\odot}$yr$^{-1}$ and filled triangles indicate $10^{-7}$\,M$_{\odot}$yr$^{-1}$. The three triangles correspond to inner disk hole sizes of 1, 2, and 4\,R$_{\odot}$.}
    \label{fig:cmd}
\end{figure}

The ground-based SMARTS data allow us to examine the color variations  during the observing period. We show the color-magnitude and near-infrared color-color diagrams in Fig.~\ref{fig:cmd} along with the extinction path by \citet{car89} indicated with black arrows assuming $R_V=3.1$. The near-infrared color-magnitude diagrams show that the pattern does not follow the extinction path in the near-infrared wavelength range but VW~Cha becomes redder as it brightens. 

We also show the publicly available ASAS-SN $g$ band measurements\footnote{https://asas-sn.osu.edu/} in Figs.~\ref{fig:lc_period} and \ref{fig:lc_period2021}. The $g$ band light curve also resembles the TESS observations. 
However, two brightening events at JD$\sim$2458605 and JD$\sim$2458642 stand out from the daily variations. Among the available optical light curves, the $g$ band observations display the largest peak-to-peak amplitude with $\sim$1.4\,mag variations in the 2019 observing season and $\sim$1.2\,mag in 2021.

\subsection{Absorption lines}
\label{sect:abs_lines}

The spectra of classical T Tauri stars encompass numerous absorption lines, however, these absorption lines are typically weaker than those of the non-accreting stars. This so-called veiling is normally caused by an excess continuum emission originating from gas at the stellar surface heated by the accretion shock. This effect is detected in the absorption lines of VW~Cha and it is further discussed in Sect. \ref{sect:veiling}.

\begin{figure}[t]
    \centering
    \includegraphics[height=0.5\textwidth, angle=90]{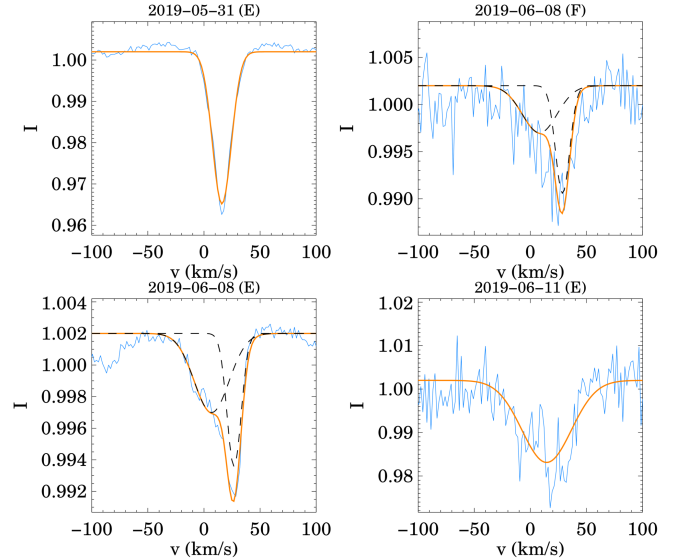}
    \caption{The LSD profiles of VW~Cha. The fitted Gaussians are indicated with the orange curves. We indicated the individual Gaussian components with dashed curves when two components were fitted.}
    \label{fig:lsd_fit}
\end{figure}

The absorption lines of VW~Cha also show some morphological variations. In order to investigate the origin of these changes, we calculated the average absorption profile using the Least Squares Deconvolution (LSD) method \citep{donati1997}. LSD is a cross-correlation technique for computing average profiles from thousands of spectral lines simultaneously. In order to achieve better signal to noise ratio, we obtained the LSD profile from the red half of the spectra ($\lambda = 584-788$~nm) using a mask with absorption lines for which the normalized flux is less than 0.985 (Fig.~\ref{fig:lsd_fit}). 

The primary component of VW~Cha is a suspected SB2, and the observed variations might originate from this close binary nature. In order to examine this possibility, we determined the radial velocity of the system by fitting Gaussians to the LSD profiles. The first ESPRESSO observation (JD=2458635.5) showed a single Gaussian profile, whereas the second ESPRESSO measurement (JD=2458642.6) hints at a split profile with two Gaussian components. The FEROS observations are noisier but as the first FEROS spectrum was taken on the same night as the second of the ESPRESSO spectrum, we applied two Gaussians for this measurement, whereas we used one component for the last epoch. We show the fitted Gaussians in Fig.~\ref{fig:lsd_fit} as well.

The results show that the average radial velocity of the system is 16.8~km/s. For the JD=2458642 night, we found two radial velocity components with a separation of $\sim$19 km/s in both the ESPRESSO and the FEROS spectra. The radial velocity results are included in Table~\ref{tab:rv}.

\begin{figure*}
    \centering
    \includegraphics[width=\textwidth]{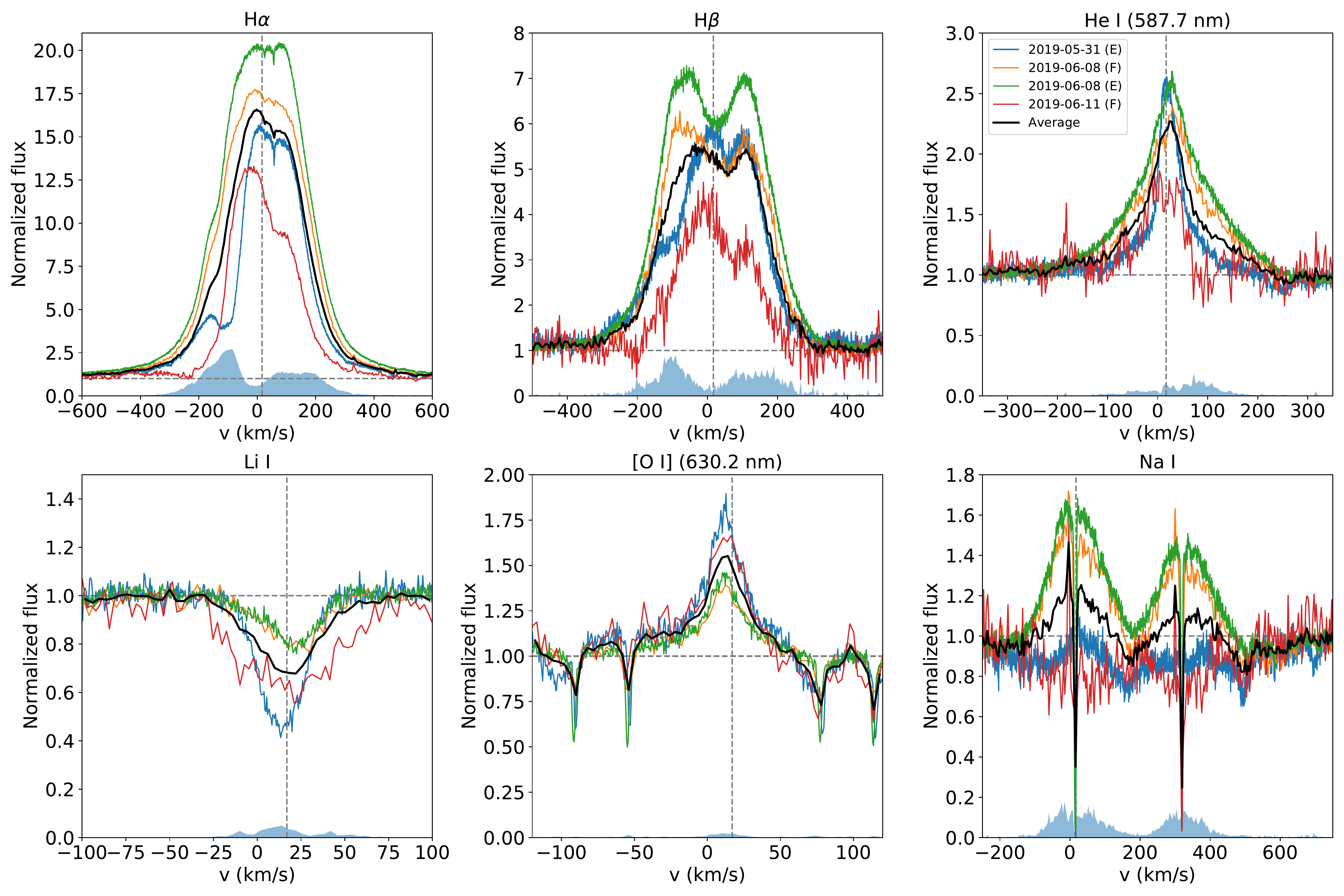}
    \caption{The normalized line profiles of the H$\alpha$, H$\beta$, He~I, Li~I, [O~I], and Na~I lines. The lines with different colors indicate the different epochs (as listed in the upper right panel), the black thick line shows the average line profile, and the blue shaded area illustrates the variance profile in each panel.}
    \label{fig:var_profile}
\end{figure*}

\subsection{Emission lines and their variability}
\label{sect:emission_lines}

The spectra of VW~Cha exhibit several emission lines (Fig.~\ref{fig:smooth}). The Balmer lines, in particular the H$\alpha$ and H$\beta$, are the most striking features of all spectra and show strong and variable line profiles. 
In addition, we identified several permitted lines, such as  He~I, Fe~I and Fe~II lines and the Na~I doublet in emission, however, not all of them were clearly detected in all epochs. 
The wavelength region of the Ca~II infrared triplet was covered only by FEROS, and we detected two of the three lines on both FEROS epochs. The 854.2\,nm line falls between two orders, therefore it was not covered. 
Moreover, [O~I] lines were also detected but we did not find any other forbidden line.

The most conspicuous feature of the spectra of VW~Cha is the H$\alpha$ emission line, which shows significant variability over the observing period (Fig. \ref{fig:var_profile}). The central peak exhibits amplitude variations which appear to be correlated with the photometric variations. When brighter photometric state was observed, the H$\alpha$ line presented stronger peak, which implies that the H$\alpha$ line became stronger by a larger factor than the continuum.
Apart from the amplitude variations, we found also morphological changes in the H$\alpha$ line. The first epoch shows a strong blueshifted absorption component around -150~km/s, which component was mostly filled by the the second epoch. This effect is also shown by the blue shaded area, which indicates the variance profile. The variance profile measures the variability in each velocity bin in the line profile, and was obtained as described in \cite{johns1995}, using the following relation:
$$\sum \limits _v = \bigg[ \frac{\sum_{i=1}^n(I_{v,i} - \overline{I}_v)^2}{(n-1)} \bigg]^{1/2},$$
where $n$ is the number of observations, $I_{v,i}$ is the intensity at a given velocity ($v$) in each observation, and $\overline{I}_v$ is the mean intensity of all the observed profiles at a given velocity $v$. 
The variance profile of the H$\alpha$ line suggests that besides the feature at $-$150~km/s, the red wing is also highly variable.

The H$\beta$ line exhibits similar amplitude variations as the H$\alpha$ line, i.e., when the system becomes brighter the emission gets stronger (see Fig.~\ref{fig:var_profile}). The line has a peculiar double peaked shape, and the relative amplitude and position of the two peaks vary over time. Our first observation indicate one peak at the systemic velocity and one redshifted peak. By the second observation, the peak at the center moved to the blueshifted side, whereas the redshifted peak remained around 130~km/s. Our last observation reveals similar peak positions as the first one, however, the redshifted peak shows more pronounced amplitude decrease than the other.

The He~I line at 587.7\,nm exhibits a peak at the rest velocity and wide wings on both sides. Besides the amplitude variations of the central peak, the wings are also changing: when brighter photometric state was observed, the line wings were more pronounced.

We also detected three [O~I] lines in all of our spectra at 557.8\,nm, 630.6\,nm, and 636.3\,nm. In Fig.~\ref{fig:var_profile}, we show 630.6\,nm line from the normalized spectra, as this was the strongest among the observed forbidden lines. The region of the [O~I] line is contaminated by telluric lines, nonetheless, we observed some amplitude variations in the normalized line profiles. However, we note that these variations appear only in the continuum normalized spectra, the line flux is almost constant during our observations. We further discuss this in Sect. \ref{sect:outflow}.
The Na I doublet shows high variability, with two observations in emission, and two measurements below the continuum (Fig.~\ref{fig:var_profile}).

We note that two measurements, the first FEROS spectrum and the second ESPRESSO spectrum, were taken at the same night only a few hours apart. The two spectra look very similar, as expected, but some strong accretion tracer lines, such as the H$\alpha$ or H$\beta$ lines, show noticeable differences (Fig.~\ref{fig:var_profile}).
In order to examine whether this discrepancy is an instrumental or a physical effect, we compared several emission and absorption lines in the two spectra. We found that most lines are in agreement within 10\%. Only the strongest lines differ $\sim$20\%, but as these are typically accretion tracers, the change might have physical origin.

\subsection{Measuring the line fluxes of the accretion tracers} 
\label{sect:acc_ines}

The H~I, He~I, Ca~II, and Na~I emission lines in the optical and near-infrared spectra of young stellar objects are tracing the accretion process \citep[e.g.][]{alc14, hartmann2016}. 
Therefore, we computed the accretion rates of VW~Cha using the fluxes of these lines detected in our data.

While FEROS spectra are Nyquist binned, the ESPRESSO spectra are oversampled. To compute the lines flux in a homogeneous way, we rebinned ESPRESSO spectra in the Nyquist way.
At first, we identified the emission lines automatically: we considered those emission lines which were above the 1$\sigma$ level of the estimated continuum level, and the line width was defined by the difference between the two line edges, where the 1$\sigma$ level intersects the line (i.e., $\lambda_{right}-\lambda_{left}$). However, due to the noise and the occasionally occurring blends with neighboring lines, we needed to manually adjust the edges of the lines in some cases, or discard the blended lines.
We determined the line flux with a python routine, which integrates the flux of the pixels contained between the local continuum, which is estimated with a linear fit, and the line. 
We computed the noise of the line by multiplying the standard deviation of the local continuum ($RMS$) by the width of the line ($\Delta \lambda_{line}$) and by the square root of the number of pixels included in the line ($N_{pix}$).
A line is considered to be detected when its $S/N \ge 3$. 
For the non detected lines, we computed the upper limit of the line flux ($F^{upp}_{line}$) by multiplying the noise by three: 
\begin{equation}
F^{upp}_{line} = 3 \times \Big( RMS \times \frac{\Delta \lambda_{line}}{R} \times \sqrt{N_{pix}} \Big)
\end{equation}
Results are presented in Tab.~\ref{tab:flux}.

\begin{deluxetable*}{llcccc}
\tablecaption{Line fluxes for the emission lines of VW~Cha \label{tab:flux}}
\tablewidth{0pt}
\tablehead{
\colhead{Species} &$\lambda_{line}$& $F^{ep1}_{obs}(E)$ & $F^{ep2}_{obs}(F)$ & $F^{ep3}_{obs}(E)$ & $F^{ep4}_{obs}(F)$ \\
\colhead{}        & [nm] &   [$10^{-14}$ erg s$^{-1}$ cm$^{-2}$]     &    [$10^{-14}$ erg s$^{-1}$ cm$^{-2}$]     &   [$10^{-14}$ erg s$^{-1}$ cm$^{-2}$]     &   [$10^{-14}$ erg s$^{-1}$ cm$^{-2}$]      }
\startdata
H3      &  656.32 & $ 382.788 \pm  0.196$ & $ 919.390 \pm  0.525$ & $ 726.983 \pm  0.236$ & $ 285.018 \pm  0.517$ \\
H4      &  486.16 & $  54.788 \pm  0.088$ & $ 122.292 \pm  0.290$ & $ 105.076 \pm  0.080$ & $  24.481 \pm  0.449$ \\
H5      &  434.07 & $  25.390 \pm  0.289$ & $  36.937 \pm  0.540$ & $  41.370 \pm  0.221$ & $   7.050 \pm  0.583$ \\
H6      &  410.2 &  $  18.400 \pm  0.344$ & $  25.994 \pm  0.946$ & $  25.940 \pm  0.280$ & $  94.672 \pm  5.548$ \\
H7      &  397.03 & $  21.356 \pm  0.263$ & $  41.300 \pm  1.145$ & $  40.656 \pm  0.205$ & $  29.485 \pm  2.256$ \\
H8      &  388.93 & $   8.756 \pm  0.348$ & $   < 4.643         $ & $  13.416 \pm  0.286$ & $   7.110 \pm  1.930$ \\
H9      &  383.56 & $  10.940 \pm  0.610$ & $ -                 $ & $  11.367 \pm  0.465$ & $ -                 $ \\
H10     &  379.81 & $   6.789 \pm  1.057$ & $ -                 $ & $  13.272 \pm  0.748$ & $ -                 $ \\
He~I/Fe~I  &  492.22 & $   2.404 \pm  0.130$ & $  14.947 \pm  0.276$ & $  10.454 \pm  0.096$ & $   2.681 \pm  0.476$ \\
He~I    &  402.64 & $   < 0.272         $ & $   < 1.687         $ & $   < 0.209         $ & $   1.057 \pm  0.315$ \\
He~I    &  447.17 & $   2.377 \pm  0.059$ & $   3.225 \pm  0.220$ & $   3.370 \pm  0.050$ & $   3.524 \pm  0.440$ \\
He~I    &  471.34 & $   0.776 \pm  0.099$ & $   0.953 \pm  0.195$ & $   0.957 \pm  0.080$ & $   1.581 \pm  0.296$ \\
He~I    &  501.6 &  $   < 0.342         $ & $  19.588 \pm  0.245$ & $   < 0.800         $ & $   4.593 \pm  0.396$ \\
He~I    &  587.6 &  $   9.293 \pm  0.098$ & $  25.047 \pm  0.206$ & $  19.603 \pm  0.088$ & $   7.345 \pm  0.275$ \\
He~I    &  667.85 & $   2.888 \pm  0.044$ & $   8.839 \pm  0.144$ & $   6.075 \pm  0.049$ & $   2.546 \pm  0.126$ \\
He~I    &  706.56 & $   2.729 \pm  0.056$ & $  12.545 \pm  0.150$ & $   6.786 \pm  0.049$ & $   7.996 \pm  0.190$ \\
He~II    &  468.61 & $   1.101 \pm  0.092$ & $   < 0.998         $ & $   1.127 \pm  0.075$ & $   < 2.063         $ \\
Ca~II (K) &  393.39 & $  19.906 \pm  0.324$ & $  40.339 \pm  1.563$ & $  46.111 \pm  0.258$ & $  21.368 \pm  2.280$ \\
Ca~II (H) &  396.87 & $  20.968 \pm  0.548$ & $  40.279 \pm  1.172$ & $  40.079 \pm  0.451$ & $  30.425 \pm  2.240$ \\
Ca~II   &  849.85 & $ -                 $ & $ 252.089 \pm  0.380$ & $ -                 $ & $  25.077 \pm  0.456$ \\
Ca~II   &  866.26 & $ -                 $ & $ 232.491 \pm  0.824$ & $ -                 $ & $  26.310 \pm  0.642$ \\
Na~I    &  589.03 & $  < 0.289          $ & $   8.470 \pm  0.445$ & $   4.650 \pm  0.160$ & $   2.190 \pm  0.284$ \\
Na~I    &  589.63 & $  < 0.291          $ & $   4.467 \pm  0.398$ & $   2.105 \pm  0.182$ & $   1.266 \pm  0.213$ \\
O~I     &  777.35 & $   7.205 \pm  0.143$ & $  29.358 \pm  0.503$ & $  18.012 \pm  0.150$ & $   5.970 \pm  0.340$ \\
O~I     &  844.68 & $ -                 $ & $  54.033 \pm  1.145$ & $ -                 $ & $  10.570 \pm  0.543$ \\
\enddata
\tablecomments{Fluxes shown here are not dereddened.}
\end{deluxetable*}

\begin{figure}[t]
    \centering
    \includegraphics[width=0.5\textwidth]{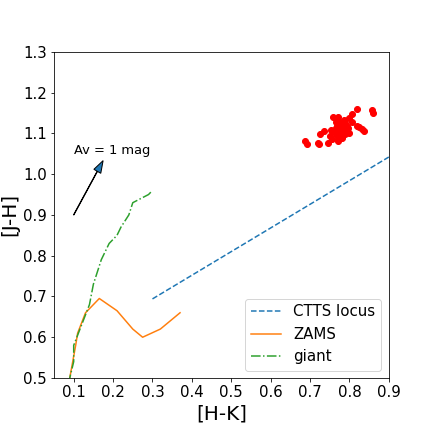}
    \caption{Color-Color diagram of VW~Cha. Red circles are data from SMARTS observations at different epochs. Blue dashed line represents the CTTS locus. The orange solid line and the green dot-dashed line correspond to the ZAMS and the giant branch, respectively. The black arrow shows the vectorial shift from the CTTS locus for a source with 1~mag of extinction.}
    \label{fig:colcol_diag}
\end{figure}

\subsection{Extinction and spectral typing} \label{sect:ext}

Pre-main sequence stars are typically effected by extinction. In order to study the stellar parameters and accretion rates, we need to estimate the extinction, which can be computed in several ways.
One of this is through the distance between the location of the star in the [$J-H$] vs. [$H-K$] diagram and the CTTS locus, the location of CTTSs if they are unextincted ($A_V=0$).  This color-color diagram is shown in Fig.~\ref{fig:colcol_diag}. 
Applying the extinction vector by \citet{car89} and assuming $R_V=3.1$, we found a mean value for the extinction of $A_V = 1.70 \pm 0.10$~mag using all our NIR photometry.

The extinction of an accreting object can also be computed from the accretion luminosity estimated by accretion tracers at different wavelengths.
We estimated the extinction from a grid of values from 0 to 10~mag, choosing the $A_V$ that minimizes the difference between the accretion luminosity derived from different lines for the same epoch. 
We used only the detected lines (see Tab.~\ref{tab:flux}). We obtained a mean value for the four epochs of $A_V = 1.54 \pm 0.27$~mag.

Lastly, we performed the spectral typing of VW~Cha by comparing our spectra with a grid of observed templates \citep[][]{man13a,man17a} from G4 to M9.5, that have a typical step of 1~spectral sub-class for spectral type G and K, and of 0.5 spectral sub-class for M~type stars, a method similar to those performed previously in the literature for single CTTSs \citep[e.g.,][]{fio21}.
In order to compare the template with the spectra of VW~Cha, we reddened the template and flux calibrated it in a window of $\Delta \lambda = 5$~nm around $\lambda = 570$~nm.  
By varying the template and the extinction, and by matching the shape and the molecular features of our data with the ones of the templates, we find that the spectral type of VW~Cha is in agreement with K5$-$K7 templates. 
We found that the mean value of the extinction for the four epochs is $A_V=1.15 \pm 0.53$~mag. 
We should point out that, for this analysis, we treated VW~Cha as it was a single star.

We note that our $A_V$ results are all in agreement within the error. In the following, we use the value provided by the [J--H] vs. [H--K] diagram, for which have the smallest error, and it is closer to the extinction values we found in the literature, for example $A_V=1.9$~mag in \citet{man17b}. 
We think the slightly different result we get from our spectra is due to the fact that we have not subtracted a slab model from the spectra to eliminate the effect of the veiling
while doing the spectral typing, and due to the different wavelength intervals analyzed, that ranges from UVB to NIR for \citet{man17b}, and which corresponds only to the optical in this work.

\begin{figure}[t]
    \centering
    \includegraphics[width=\columnwidth]{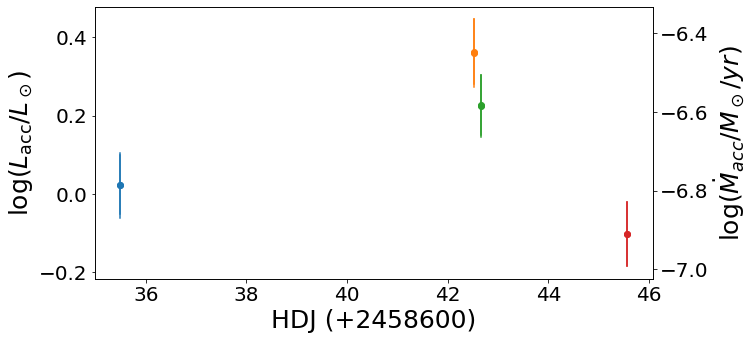}
    \caption{Accretion luminosity and mass accretion rate as a function of the time.}
    \label{fig:lacc_time}
\end{figure}

\begin{deluxetable}{lcc}
\tablecaption{Accretion parameters for VW~Cha \label{tab:accretion}}
\tablehead{
\colhead{Date} & \colhead{$\log \lacc$} &\colhead{$\log \macc$} \\
\colhead{ } & \colhead{[$\lsun$]} &\colhead{[$\msun$ yr$^{-1}$]} 
}
\startdata
2019 May 31 (E) & $0.022 \pm 0.083$ & $-6.785 \pm 0.075$\\
2019 June 08 (F) & $0.360 \pm 0.086$ & $-6.447 \pm 0.082$\\
2019 June 08 (E) & $0.225 \pm 0.079$ & $-6.582 \pm 0.075$\\
2019 June 11 (F) & $-0.103 \pm 0.082$& $-6.910 \pm 0.082$
\enddata
\tablecomments{The values presented here are without taking into account the error on stellar parameters.}
\end{deluxetable}

\subsection{Accretion Luminosity and accretion rate} \label{sect:acc}

We used the detected accretion tracer lines from Tab.~\ref{tab:flux} to estimate the accretion luminosity ($\lacc$) of the VW~Cha system. 
The accretion luminosity is related to the line luminosity of accretion tracers ($L_{line}$) by empirical relations. 
We computed the line luminosity as $L_{\rm line}=4\pi d^2 F_{\rm line}$, where $F_{\rm line}$ is the extinction corrected line flux and $d=194.5$~pc is the distance \citep[][]{bailerjones2021}.
Then, we estimated the accretion luminosity for each line by using the relations from \citet{alc17}:
 \begin{equation}
 \log_{10} \left( \frac{\lacc}{\lsun}\right) = a \log_{10}\left( \frac{L_{\rm line}}{\lsun} \right) + b
 \label{eqLacc}
 \end{equation}
where $a$ and $b$ vary from line to line. 
The relations in \citep{alc17} are based on spectroscopic observations of a sample of young stellar objects. They have some natural uncertainty originating from the measurement of the line flux, and the linear fit between the line luminosity and the accretion luminosity. The overall uncertainties for the relations differ from line to line, and the standard deviation of the liner fit range from $\sigma$=0.26 to $\sigma$=0.45 among those relations which are suggested for deriving the accretion luminosity.
The error on each line is obtained by propagating the error of $F_{\rm line}$. 
As the best estimate of the accretion luminosity for each epoch, we used the mean value of all the accretion luminosity provided by the detected lines of that epoch. The error on the mean accretion luminosity for each epoch was computed by dividing the standard deviation of the errors by the square root of the number of lines used. Results are shown in Tab.~\ref{tab:accretion}.

We computed the mass accretion rate with the following equation:
\begin{equation}
 \label{eqmacc}
  \macc \sim \left(1 - \frac{\rstar}{R_{\rm in}}\right)^{-1} \frac{\lacc \rstar}{G \mstar}
\end{equation}
where $R_{\rm in}$ is the inner-disk radius which we assume to be $R_{\rm in} \sim 5 R_\star$ \citep{hartmann1998}, $\mstar = 0.75 \pm ^{0.5}_{0.35}$~$\msun$, and $\rstar = 2.55 \pm 0.29$~$\rsun$ from \citet[][]{dae13}.
We computed the error on the mass accretion rate in the same way we have done for the accretion luminosity, and not taking into account the uncertainty on the stellar parameters. Results are shown in Tab.~\ref{tab:accretion}.
Fig.~\ref{fig:lacc_time} displays the accretion luminosity and the mass accretion rate as a function of the time, which show increased accretion luminosity and accretion rate in the second and third epoch, and a decrease by the last epoch.

\begin{figure}[t]
    \centering
    \includegraphics[width=0.5\textwidth] {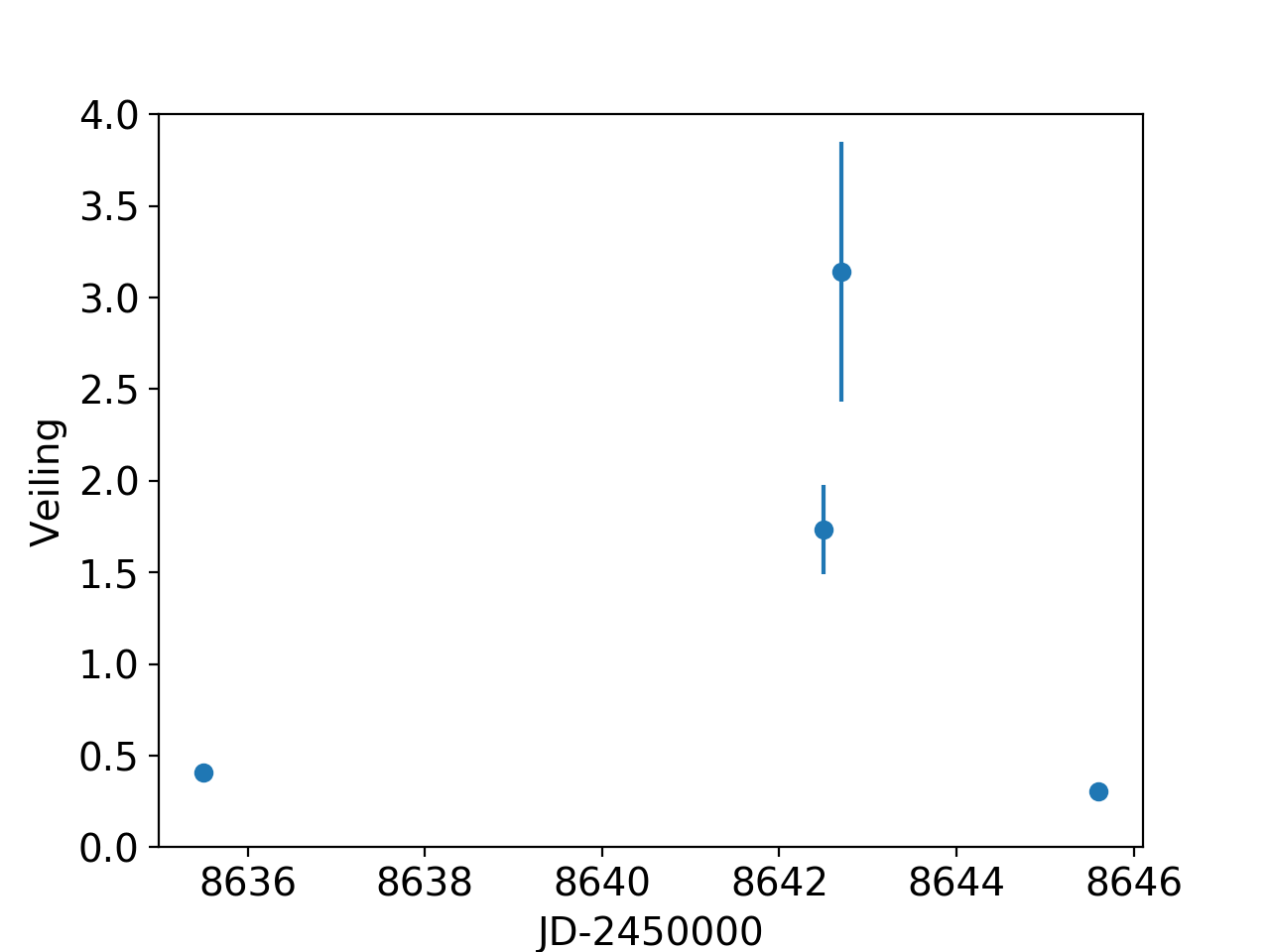}
    \caption{Absolute veiling measurements.}
    \label{fig:veiling}
\end{figure}

\subsection{Veiling}
\label{sect:veiling}

The absorption spectra of classical T Tauri stars are often veiled by an additional continuum emission due to accretion. As the veiling is also indicative of the accretion rate, we expect that when the accretion rate changes, the veiling and the stellar brightness changes accordingly. We calculated the veiling ($V$) for our four spectroscopic observations using the following relation:
$$W^{eq}_0 = W^{eq}(V+1),$$
where $W^{eq}$ and $W^{eq}_0$ is the measured equivalent width of the absorption line of the target and a reference object, respectively.

Instead of using one absorption line, we used the LSD profiles obtained from the spectra previously. For reference, we obtained the LSD profile based on the same criteria for a non-accreting K7-type weak-line T~Tauri star, TWA 6, whose spectrum is available from the Polarbase database\footnote{http://polarbase.irap.omp.eu/}. We used this reference in order to calculate the absolute veiling. The results are shown in Fig.~\ref{fig:veiling}.

It must be noted that the veiling is expected to be wavelength dependent in a way that typically larger veiling is measured at shorter wavelengths \citep[e.g.,][]{dodin2013}.
The use of the LSD profile ignores this wavelength dependence, as it serves as an average absorption line profile.
In order to test the significance of its effect, we selected a few individual absorption  lines above 500\,nm to avoid the noisiest part of the spectra, and calculated the veiling using them.
Our results show that, within the uncertainties, the general temporal trend is similar to the bulk measurement provided by the use of the LSD profiles.
However, the measurements using individual lines did not reproduce the expected trend (i.e., larger veiling at shorter wavelength), but instead resulted in a scatter around the values determined by using the LSD profile.
This might be due to the fact that in many cases, the individual absorption lines were still noisy and weak, furthermore, the stronger ones were often blended with neighboring lines.

\newpage

\section{Discussion} \label{sect:discussion}

\subsection{Near-infrared color variations}
\label{sect:color}
Our ground-based near-infrared observations of VW~Cha reveal an unusual trend: the source becomes redder as it brightens (Fig.~\ref{fig:cmd}). In order to examine the physical origin of this behavior, we compared our results with three models described by \cite{carpenter2001} which use the models by \cite{meyer1997}.

The first model involves starspots, which often appear on the surface of T~Tauri stars, and can be responsible for variability observed in these systems. \cite{carpenter2001} inspected the impact of starspots at near-infrared wavelengths, and found that cool spots with small fractional coverage result in nearly colorless fluctuations. In case of larger spot coverages and hot spots, the color-magnitude diagram shows that the system becomes bluer as it gets brighter.

Inhomogeneities in the inner circumstellar environment can cause changing extinction, which also results in photometric variability. 
The extinction model introduces a positive slope in the near-infrared color-magnitude diagrams, i.e., the object gets redder as it becomes fainter. In order to compare our observations with this model, we indicated the extinction arrow in Fig.~\ref{fig:cmd} corresponding to $A_V=1$\,mag, which shows that our data are not consistent with changing extinction.

The disk model discussed by \cite{carpenter2001} and \cite{meyer1997} attributes the near-infrared variation to changes in the accretion disk, such as variable inner disk structure or accretion rate. 
This model results in a negative slope in the near-infrared color-magnitude diagrams (i.e., the source gets redder as it brightens) and predicts a shallower slope in the [$J-H$] vs. [$H-K_S$] diagram than the above mentioned spot and extinction models. 
In order to compare the disk model with our observations, we indicated the color changes predicted by this model compared to our bluest data point  with triangles in Fig.~\ref{fig:cmd}. 
The open triangles represent mass accretion rate of $10^{-8.5}$\,M$_{\odot}$yr$^{-1}$, and the filled triangles show $10^{-7.0}$\,M$_{\odot}$yr$^{-1}$ \citep{carpenter2001}. The three triangles correspond to disk inner hole sizes of 1, 2, and 4\,R$_\odot$.
These models of \cite{meyer1997} include a fiducial M0 type star with fixed $\mstar=0.5\,\msun$, $\rstar = 1.8\,\rsun$, and $\teff = 4000$\,K. These stellar parameters describe a typical T~Tauri star, but do not agree entirely with the parameters of VW~Cha, therefore, our analysis analysis give a qualitative explanation for the observed trajectory.
The comparison suggests that the near-infrared color variations seen in our dataset are orthogonal to the trajectory of the spot and the extinction models, and they are consistent with those predicted by the disk model. This means that the observed color variations originate from the changes in the accretion disk.

We also considered an other set of models by \cite{dalessio1998, dalessio1999, dalessio2001}, which provides the spectral energy distributions (SED) for accretion disks with various disk parameters\footnote{https://lweb.cfa.harvard.edu/youngstars/dalessio/} ($\rdisk$, $\rhole$, $\macc$, $\alpha$, $i$) for systems with different central stars.
First, we selected a central star with stellar parameters closest to our target ($\teff$=4000\,K, 1\,Myr, $\rstar=2.64\,\rsun$, $\mstar = 0.7\,\msun$, $\lstar = 1.6\,\lsun$), and we studied the corresponding disk models.
From the available models, we chose the ones with $\rdisk = 100\,\rm au$, $i = 30^{\circ}$, and $\alpha=0.01$, which parameters are expected for VW~Cha.
This left two parameters free: $\rhole$ and $\macc$.
Unfortunately, the $\rhole$ and $\macc$ grid of these models are sparser than the ones presented by \cite{meyer1997}, therefore, these models are less representative of VW~Cha.
Nonetheless, we attempted to compare our data and the models by placing our observations on the SEDs of the models with the following pairs of parameters: $\rhole = 9\,\rstar$ and $\macc = 10^{-9}\,\msun/\rm yr$,  $\rhole = 9\,\rstar$ and $\macc = 10^{-8}\,\msun/\rm yr$, $\rhole = 11\,\rstar$ and $\macc = 10^{-7}\,\msun/\rm yr$, $\rhole = 22\,\rstar$ and $\macc = 10^{-6}\,\msun/\rm yr$.
This comparison did not lead to any strong conclusions, which might arise from the fact that both the inner disk hole size and the accretion rate varies from model to model, and there is no option to change only one of these parameters.
Additionally, we converted the fluxes close to the $JHK$ bands from the SED models to magnitudes and we constructed the NIR color-magnitude and color-color diagrams. 
In general, we found that these models also reproduce the observed trend (i.e., the source becomes redder when it brightens).
However, the trend appears in the opposite way compared to the \cite{carpenter2001} models: in \cite{carpenter2001}, larger hole sizes correspond to smaller infrared excess, whereas the \cite{dalessio1998} models resulted in an opposite trend, i. e., larger hole sizes correspond to larger infrared excess.
Again, this might arise from the fact that for the \cite{dalessio1998} models both the inner disk hole size and the accretion rate change from model to model (larger inner hole size is paired with larger accretion rate). 
This does not contradict the \cite{carpenter2001} models: their models with larger accretion rate (filled triangles) also predict redder colors.

Apart from changes in the inner disk hole size, other changes in the inner disk structure may contribute to the observed trend in the NIR color-magnitude and color-color diagrams. These involve e.g. variations in the thickness of the inner disk edge of a warped disk \citep{mahdavi1998, lai1999, terquem2000}, or the spectroscopic binary nature (see Sect.~\ref{sect:sb2}) might also disturb the inner disk edge and cause structural changes.

\subsection{Changes between the brighter and fainter states}
The light curves of VW~Cha display significant photometric changes: we observed 0.6$-$0.9\,mag peak-to-peak variability in the different photometric bands in 2019. This dynamical range is dominated by the brightening event at JD$\sim$2458642. We measured a spectrum before, during and after this event, which allows us to study its nature and origin.

The spectra taken at the bright and the faint states show noteworthy differences. The Balmer lines undergo the most striking evolution, however, other accretion tracers also vary in time. The significant amplitude variations of these lines in the normalized spectra suggest that accretion rate change might be the major contributor of the observed effect. The computed line fluxes also support this concept, since we measured higher line fluxes in the brighter state. The veiling measurements (Fig.~\ref{fig:veiling}) are also consistent with the described picture, i.e., the veiling was higher during the brighter state, indicating higher accretion rate.
Indeed, we calculated the accretion rates for each epoch, which show an increase by a factor of 2.17 by the brighter state (Tab.~\ref{tab:accretion}).

The emission lines exhibit morphological evolution as well during the previously discussed brightening event. The He~I lines show only mild changes, however, the wings become more pronounced in the brighter epoch. In addition, the variance profile highlights a slight asymmetry: the red wing varies more during our observations than the blue wing.
This broad component extends to $\sim$250\,km/s, and is thought to form in the magnetospheric flow \citep{hartmann2016}. 
The more pronounced broad component in the brighter state might arise due to the increased accretion.
The H$\beta$ line exhibits an interesting double peaked profile where both the strength and the position of the two components are changing.

The H$\alpha$ line has the most striking morphological differences between the different measurements. The first epoch shows an emission peak with a strong blueshifted absorption component. This suggests that the accretion is accompanied by an outflow, which is discussed in Sect.~\ref{sect:outflow}. As the accretion rate increases by the second observation, this blueshifted absorption component becomes filled.

\cite{stauffer2014} examined the light curves and spectra for a sample of young stellar objects in NGC~2264, including several CTTSs. They studied light curves dominated by accretion bursts, similarly to the that of VW~Cha, and found that these brief – several hours to a day-long - brightenings have amplitudes in the range of 5\%$-$50\% of the quiescent level. The longer duration (several days) of the brightening event at JD$\sim$2458642 in the light curve of VW~Cha and the fine structure, revealed by the high cadence TESS data, suggests that this event might be a superposition of multiple smaller accretion bursts. The spectral features that we found are also consistent with the results of \cite{stauffer2014}, i.e., accretion burst stars have modestly structured and centrally peaked H$\alpha$ line during the accretion burst, and the 667.8\,nm He~I line in emission.

\begin{figure}[ht!]
    \centering
    \includegraphics[width=\columnwidth]{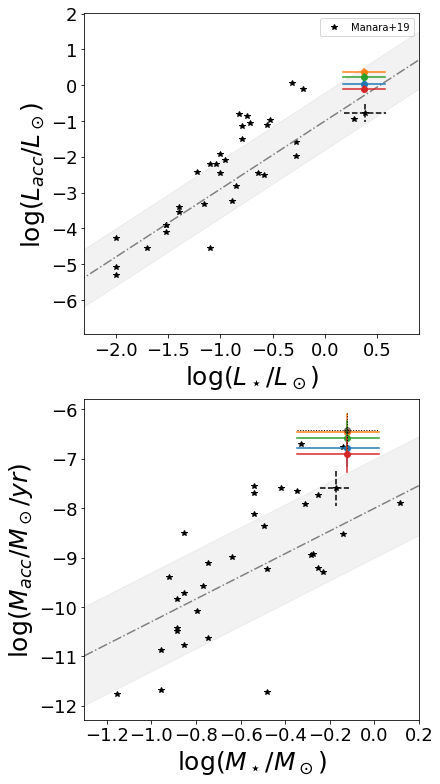}
    \caption{{\it Top}: Accretion luminosity as a function of stellar luminosity. {\it Bottom}: Mass accretion rate as a function of the stellar mass. Black stars are results of the Chamaeleon~I survey of CTTSs from \citet[][]{man19}. For this work, we plot with black dashed lines the error bars only for VW~Cha. Black filled circle corresponds to the mass accretion rate computed by using the accretion luminosity from \citet[][]{man19} and stellar parameters from \citet[][]{dae13}, errors bars are shown with dotted black lines. The linear fit is taken from \citet[][]{fio21}. Colored filled circles show the results of this work, different colors correspond to different epochs as in Fig.~\ref{fig:lacc_time}. }
    \label{fig:lacc_lstar}
\end{figure}

In Fig.~\ref{fig:lacc_lstar}, we compare VW~Cha with other sources from the Chamaeleon~I star forming region using the results of the accretion rates and the stellar parameters of \cite{man19}, and we used the fit from \cite{fio21}. We depict the accretion luminosity against the stellar luminosity on the top panel and the accretion rate against the stellar mass in the bottom panel. Our results show that, as we expected based on previous studies, VW~Cha has high accretion luminosity and accretion rate compared to the other members of this sample, and places VW~Cha at the high end of the distribution. While our new results match perfectly with the fitted relation for Cha~I in the $\lacc$ vs. $\lstar$ diagram, we note that the $\macc$ is larger than the trend fitted for the overall region.

\subsection{Outflow}
\label{sect:outflow}
Mass ejection is often observed in CTTSs. Strong winds and outflows are thought to be accretion powered, and they lead to forbidden line emission and often to blueshifted absorption components superimposed on background emission \citep{calvet1997, bally2016}.

VW~Cha shows signs of outflow in the form of forbidden [O I] emission lines and blueshifted absorption component of the H$\alpha$ line. This absorption component, however, is not always present. The reason for this is the fact that it is superimposed on background emission arising due to the magnetospheric accretion. When the accretion rate increases, the accretion powered emission component strengthens and fills the absorption component. 

The [O~I] emission lines are also detected in the spectra at 557.7\,nm, at 630.0\,nm and 636.3\,nm. 
They appear to be variable based on the normalized line profile (Fig.~\ref{fig:var_profile} middle panel in bottom row), however, this line is expected to be intrinsically stable on a time scale of days \citep{petrov2011}, and the observed effect might be due to the fact that during the brighter state, the continuum level rose. Indeed, we calculated the line fluxes of the detected forbidden oxygen lines, which differ only a few percent suggesting that the normalized line profiles change due to continuum variations.

The observed line profile carries information on the kinematics of the emitting gas. In the case of VW~Cha, the [O~I] exhibits one, low velocity component (FWHM\,$\sim 30 - 60$~km/s), which is slightly blue shifted compared to the systemic velocity. 
The low-velocity component with peak velocity $v_p\lesssim$~30~km~s$^{-1}$ is thought to be associated with extended disk winds \citep{kwan1995, pascucci2020}.  For a sample of T~Tauri stars, \cite{gangi2020} examined the component at the lowest peak velocity of the [O~I] line, and our results for VW~Cha is consistent with the typical line kinematic parameters that \cite{gangi2020} found, suggesting that it originates from the innermost few au of the system.

\begin{deluxetable}{cccc}
\tablecaption{Radial velocity measurements for VW~Cha. N12 indicates the results from \cite{nguyen2012}.
For the epochs we report only one value, we were not able to resolve the two components. \label{tab:rv}}
\tablewidth{0pt}
\tablehead{
\colhead{JD$-$2450000} & \colhead{RV$_1$} & \colhead{RV$_2$} & \colhead{Ref.}
}
\startdata
3789.26999	& 23.72 $\pm$ 0.29 & 5.79 $\pm$ 0.41 & N12 \\
3836.16847	& 27.16 $\pm$ 0.21 & 6.02 $\pm$ 0.37 & N12 \\
4072.37156	& 26.31 $\pm$ 0.26 & 2.51 $\pm$ 0.35 & N12 \\
8635.49998 & \multicolumn{2}{c}{16.00$\pm$ 0.10} & This work \\
8642.52353 & 28.66$\pm$0.43 & 8.26$\pm$0.20 & This work \\
8642.66358 & 26.82$\pm$0.42 & 6.83$\pm$3.26 & This work \\
8645.57175 & \multicolumn{2}{c}{14.49$\pm$0.14} & This work \\
\enddata
\end{deluxetable}

\subsection{Spectroscopic binary}
\label{sect:sb2}

The primary component of the VW~Cha system is a suspected spectroscopic binary (SB2). \cite{melo2003} observed line doubling based on their cross correlation function, and report 15.31\,km/s as a mean radial velocity for the system. \cite{nguyen2012} found evidence for two components of the primary separated by 20\,km/s. 

We determined the radial velocities by fitting the LSD profiles, and reported the results in Table~\ref{tab:rv} along with the radial velocities from \cite{nguyen2012}. 
In the cases where we were able to resolve the two components, our results are in agreement with \cite{nguyen2012}, 
where they find that their spectra of the primary show evidence for two components (Aa, Ab) separated by 20 km/s (see their Fig.~10.10). As in our observations, we found one component in the first epoch, two components in the second and third epochs, and again one component in the fourth epoch ten days later, we speculate an orbital period for the SB2 of $\sim$10\,days.
In case the primary is indeed an SB2, and we need to account for the possibility that both components contribute to the observed spectroscopic variations in our new measurements. 
\cite{frasca2021} examined a similar multiple system, CVSO~104, where they confirmed that the target is a double-lined spectroscopic binary. They determined the mass accretion rates of the two components from the fluxes of the lines where the profiles of the two components could be deblended. In our case, the individual components of the accretion tracers, such as the H$\alpha$ or the He~I lines appear to be blended (see Fig.~\ref{fig:var_profile}), similarly to the H$\alpha$ line presented in \cite{frasca2021}, therefore we were not able to determine the accretion rates for the individual components. The H$\beta$ is the only line which shows a double peaked line profile, however, the peak velocities do not correspond to the radial velocities that we determined for each epoch.

\section{Summary}

We designed a complex observing campaign for the multiple T~Tauri system VW~Cha with multifilter photometry, and high-resolution optical spectroscopy contemporaneously with the TESS space telescope observations.
The high cadence TESS photometry reveals variations on hourly-daily timescales with a peak-to-peak amplitude of $\sim$0.8\,mag.
The near-infrared data show that the amplitude of the variability is slightly increasing towards the longer wavelengths. The NIR color-magnitude diagrams show the unusual trend of the source becoming redder as it brightens. This trend cannot be explained with stellar spots or variations in the extinction, however, it can be associated with changes in the accretion disk. 

The light curves show a distinct brightening event in the second half of our observing period at all wavelengths. 
We took spectra before, during, and after this event allowing us to examine its nature. 
Our results suggest that it originates from increased accretion. 
The accretion tracers, the measured mass accretion rate, and the veiling is also consistent with this scenario.

Apart from the accretion, the spectra indicate the presence of an outflow as well. We found [O~I] lines with one, low velocity component, which is considered to be associated with disk winds.

Moreover, VW~Cha is known to be a multiple system, however, some studies suggest that the primary itself is a spectroscopic binary. Our observations also hint at the spectroscopic binary nature with a speculated orbital period around 10\,days.

\acknowledgments

This project has received funding from the European Research Council (ERC) under the European Union's Horizon 2020 research and innovation programme under grant agreement No 716155 (SACCRED). 
G.Zs. is supported by the \'UNKP20-3 New National Excellence Program of the Ministry for Innovation and Technology from the source of the National Research, Development and Innovation Fund.
The research leading to these results has received funding from the LP2018-7 Lend\"ulet grants of the Hungarian Academy of Sciences.
Based on observations collected at the European Southern Observatory under ESO programmes 2103.C-5025, 0103.A-9008, and 0100.C-0708(A).
This work has received  financial support of the Hungarian National Research, Development and Innovation Office – NKFIH Grant K-138962.
This project has received funding from the European Union's Horizon 2020 research and innovation programme under the Marie Sklodowska-Curie grant agreement No 823823 (DUSTBUSTERS).
This work was partly funded by the Deutsche Forschungsgemeinschaft (DFG, German Research Foundation) - 325594231. 
This research received financial support from the project PRIN-INAF 2019 "Spectroscopically Tracing the Disk Dispersal Evolution"
%

\vspace{5mm}
\facilities{TESS, SMARTS, ESPRESSO, FEROS}


\software{astropy \citep{2013A&A...558A..33A}  
          }



\clearpage
\appendix

\section{Ground-based photometry}
We present in Table \ref{tab:photometry} our optical and near-infrared SMARTS photometry.

\begin{deluxetable*}{ccccccc}[ht!]
\tablecaption{\label{tab:photometry} Optical and near-infrared photometry of VW~Cha.}
\tablewidth{0pt}
\tablehead{
\colhead{JD$-$2450000} & $V\pm \Delta V$ & $R_C \pm \Delta R_C$ & $I_C\pm \Delta I_C$ & $J \pm \Delta J$ &  $H \pm \Delta H$ & $K \pm \Delta K$
}
\decimalcolnumbers
\startdata
8603.56 & 12.688 $\pm$ 0.038 & 11.928 $\pm$ 0.020 & 10.884 $\pm$ 0.005  & 8.754 $\pm$ 0.042 & 7.617 $\pm$ 0.018 & 6.818 $\pm$ 0.012 \\
8604.56 & 12.567 $\pm$ 0.010 & 11.836 $\pm$ 0.009 & 10.829 $\pm$ 0.001  & 8.664 $\pm$ 0.009 & 7.545 $\pm$ 0.004 & 6.761 $\pm$ 0.009 \\
8605.57 & 12.447 $\pm$ 0.015 & 11.827 $\pm$ 0.006 & 10.865 $\pm$ 0.001  & 8.718 $\pm$ 0.003 & 7.629 $\pm$ 0.005 & 6.848 $\pm$ 0.007 \\
8606.53 &                    &                    & 11.021 $\pm$ 0.003  & 8.845 $\pm$ 0.008 & 7.727 $\pm$ 0.005 & 6.953 $\pm$ 0.004 \\
8607.52 &                    &                    & 10.994 $\pm$ 0.001  & 8.805 $\pm$ 0.009 & 7.700 $\pm$ 0.009 & 6.965 $\pm$ 0.009 \\
8608.54 &                    &                    & 10.991 $\pm$ 0.001  & 8.744 $\pm$ 0.010 & 7.635 $\pm$ 0.006 & 6.882 $\pm$ 0.008 \\
8609.52 &                    &                    & 10.783 $\pm$ 0.001  & 8.636 $\pm$ 0.002 & 7.528 $\pm$ 0.007 & 6.755 $\pm$ 0.007 \\
8611.52 &                    &                    & 10.768 $\pm$ 0.003  & 8.665 $\pm$ 0.010 & 7.564 $\pm$ 0.012 & 6.794 $\pm$ 0.007 \\
8612.54 &                    &                    & 10.763 $\pm$ 0.002  & 8.692 $\pm$ 0.011 & 7.598 $\pm$ 0.004 & 6.847 $\pm$ 0.006 \\
8613.51 &                    &                    & 10.720 $\pm$ 0.002  & 8.661 $\pm$ 0.006 & 7.558 $\pm$ 0.010 & 6.782 $\pm$ 0.003 \\
8614.54 &                    &                    & 10.790 $\pm$ 0.003  & 8.611 $\pm$ 0.009 & 7.512 $\pm$ 0.008 & 6.723 $\pm$ 0.010 \\
8615.52 &                    &                    & 10.719 $\pm$ 0.002  & 8.578 $\pm$ 0.008 & 7.444 $\pm$ 0.001 & 6.657 $\pm$ 0.010 \\
8616.52 &                    &                    & 10.841 $\pm$ 0.003  & 8.696 $\pm$ 0.010 & 7.577 $\pm$ 0.011 & 6.758 $\pm$ 0.012 \\
8617.54 &                    &                    & 10.809 $\pm$ 0.002  & 8.657 $\pm$ 0.005 & 7.547 $\pm$ 0.005 & 6.762 $\pm$ 0.003 \\
8618.54 &                    &                    & 10.896 $\pm$ 0.003  & 8.729 $\pm$ 0.009 & 7.626 $\pm$ 0.009 & 6.859 $\pm$ 0.001 \\
8619.51 &                    &                    & 10.754 $\pm$ 0.064  & 8.786 $\pm$ 0.004 & 7.685 $\pm$ 0.009 & 6.886 $\pm$ 0.008 \\
8620.51 &                    &                    & 10.899 $\pm$ 0.002  & 8.779 $\pm$ 0.013 & 7.683 $\pm$ 0.013 & 6.914 $\pm$ 0.006 \\
8621.54 &                    &                    & 10.779 $\pm$ 0.002  & 8.673 $\pm$ 0.005 & 7.547 $\pm$ 0.012 & 6.757 $\pm$ 0.021 \\
8622.51 &                    &                    & 10.834 $\pm$ 0.002  & 8.682 $\pm$ 0.012 & 7.542 $\pm$ 0.014 & 6.783 $\pm$ 0.014 \\
8623.53 &                    &                    & 10.928 $\pm$ 0.022  & 8.535 $\pm$ 0.029 & 7.673 $\pm$ 0.074 & 6.678 $\pm$ 0.043 \\
8624.51 &                    &                    & 10.934 $\pm$ 0.004  & 8.736 $\pm$ 0.010 & 7.646 $\pm$ 0.014 & 6.874 $\pm$ 0.001 \\
8625.52 &                    &                    & 11.133 $\pm$ 0.005  & 8.914 $\pm$ 0.004 & 7.838 $\pm$ 0.008 & 7.119 $\pm$ 0.011 \\
8626.52 &                    &                    & 11.158 $\pm$ 0.003  & 8.919 $\pm$ 0.006 & 7.845 $\pm$ 0.015 & 7.153 $\pm$ 0.013 \\
8627.52 &                    &                    & 11.095 $\pm$ 0.002  & 8.846 $\pm$ 0.002 & 7.765 $\pm$ 0.006 & 7.077 $\pm$ 0.003 \\
8630.50 &                    &                    & …                   & 8.746 $\pm$ 0.001 & …                 & …                 \\
8635.66 &                    &                    & 10.915 $\pm$ 0.003  & 8.665 $\pm$ 0.010 & 7.559 $\pm$ 0.005 & 6.801 $\pm$ 0.010 \\
8636.53 &                    &                    & 10.822 $\pm$ 0.002  & 8.577 $\pm$ 0.005 & 7.475 $\pm$ 0.005 & 6.696 $\pm$ 0.001 \\
8637.51 &                    &                    & 10.732 $\pm$ 0.001  & 8.544 $\pm$ 0.005 & 7.415 $\pm$ 0.010 & 6.638 $\pm$ 0.010 \\
8638.49 &                    &                    & 10.921 $\pm$ 0.001  & 8.712 $\pm$ 0.003 & 7.583 $\pm$ 0.007 & 6.818 $\pm$ 0.023 \\
8639.54 &                    &                    & 10.913 $\pm$ 0.004  & 8.632 $\pm$ 0.011 & 7.473 $\pm$ 0.003 & 6.653 $\pm$ 0.012 \\
8640.52 &                    &                    & 10.762 $\pm$ 0.001  & 8.561 $\pm$ 0.003 & 7.412 $\pm$ 0.009 & 6.553 $\pm$ 0.006 \\
8641.50 &                    &                    & 10.511 $\pm$ 0.003  & 8.334 $\pm$ 0.003 & 7.177 $\pm$ 0.003 & 6.319 $\pm$ 0.007 \\
8642.52 &                    &                    & 10.465 $\pm$ 0.003  & 8.359 $\pm$ 0.007 & 7.211 $\pm$ 0.003 & 6.404 $\pm$ 0.007 \\
8643.49 &                    &                    & 10.463 $\pm$ 0.002  & 8.414 $\pm$ 0.005 & 7.308 $\pm$ 0.004 & 6.470 $\pm$ 0.005 \\
8644.51 &                    &                    & 10.798 $\pm$ 0.001  & 8.723 $\pm$ 0.005 & 7.646 $\pm$ 0.013 & 6.901 $\pm$ 0.005 \\
8645.49 &                    &                    & 10.968 $\pm$ 0.002  & 8.817 $\pm$ 0.005 & 7.743 $\pm$ 0.006 & 7.021 $\pm$ 0.015
\enddata
\end{deluxetable*}


\clearpage
\bibliography{sample63}{}
\bibliographystyle{aasjournal}



\end{document}